\documentclass[a4paper,11pt]{article}
\usepackage[reqno,fleqn]{amsmath}
\usepackage{color,graphicx}
\usepackage[T1]{fontenc}
\author{Behnam Mohammadi$^a$\footnote{be.mohammadi@urmia.ac.ir}, Elnaz Amirkhanlou$^b$\footnote{eliamirkhanlou@yahoo.com}\\
Department of Physics, Urmia University, Urmia, Iran}
\title{Calculation of branching fraction and $CP$ violation in $B^-\rightarrow D_s^-D^0$ decay}
\begin{document}
\maketitle
\begin{abstract}

The most precise measurement of the $CP$ asymmetry in the decay
$B^-\rightarrow D_s^-D^0$ has been reported by LHCb collaboration
with the value of $(-0.4\pm0.5\pm0.5)\%$. In this study, the $CP$
violation in the decay $B^-\rightarrow D_s^-D^0$ has been
calculated under the factorization approach. This decay mode
includes current-current tree and penguin diagrams and their
amplitudes are considered separately. In each of the tree and
penguin amplitudes, the strong and weak phases have been
introduced. The $CP$ asymmetry has been calculated in this work to
be $(-0.35\pm0.03)\%$. Finally, from the sum of the amplitudes, we
have calculated the total amplitude and obtained comparable
results with experimental value for  the branching ratio of
$B^-\rightarrow D_s^-D^0$ decay.
\end{abstract}

\section{Introduction}
In the Standard Model (SM), the primary importance of studying the
nonleptonic two-body decays of $B$ mesons is to explore $CP$
violation and flavour parameters. The different weak and strong
interaction phases that arise from the interference of several
competing amplitudes play an important role in $CP$ violation. The
weak complex phases are obtained from the argument of quark mixing
matrix elements entering each amplitude vertex, which follow the
unitarity triangle relation,
$V_{ub}^*V_{uq}+V_{cb}^*V_{cq}+V_{tb}^*V_{tq}=0$ $(q=d,s)$. All
measured $CP$ asymmetries are in the SM related through this
unitarity condition. The weak phases are introduced in a standard
convention as $\phi_1=-{\rm{arg}}(V_{tq})$ and
$\phi_2={\rm{arg}}(V^*_{ub})$, the strong interaction phase
differences between tree and penguin amplitudes, $\delta$, is
another phase that is necessary for $CP$ violation to occur.\\
In some charm decays of $B$ mesons, such as the $B^-\rightarrow
D_s^-D^0$ decay, the decay rates are obtained for combined
particles and antiparticles, while such decays are affected $CP$
violation. It is in such cases
that the weak and strong phases induce the violation that has occurred for the charge and parity in reality.\\
The LHCb collaboration has reported the first measurement of $CP$
asymmetry in the $B^-\rightarrow D_s^-D^0$ decay \cite{LHCb2}. The
value was measured to be $(-0.4\pm0.5\pm0.5)\%$. There are also
several experimental results and their averages for the
$CP$-averaged branching ratio of the $B^-\rightarrow D_s^-D^0$
decay, some of them are shown in
Tab.\ref{tab1}.\\
\begin{table}[t]
\centering\caption{\label{tab1} Some experimental results and
their averages of $CP$-averaged branching ratio for
$B^-\rightarrow D_s^-D^0$ decay (in units of $10^{-3}$).}
\begin{tabular}{|c||c|}
  \hline
  Refs. & $\mathcal{B}(B^-\rightarrow D_s^-D^0)$ \\
  \hline\hline
 BABAR \cite{BABAR1} & $13.3\pm1.8\pm3.2$ \\
   \hline
 Belle \cite{Belle1} & $9.5\pm0.2$ \\
  \hline
 LHCb \cite{LHCb1} & $8.6\pm0.2\pm0.4\pm1.0$ \\
  \hline
  HFLAV \cite{HFAG1} & $13.3\pm3.7$ \\
   \hline
 PDG(2020) Avg. \cite{PDG1}& $9.0\pm0.9$ \\
   \hline
\end{tabular}
\end{table}
In this work, we have calculated the $CP$ violation and
$CP$-averaged branching ratio for the $B^-\rightarrow D_s^-D^0$
decay. First, we have drawn all the contributions of decay
diagrams in accordance with Feynman rules. Using the factorization
approaches, the matrix elements of effective Hamiltonian have been
evaluated. In these approach, the simple factorization of the
elements of the hadron matrix appears as the product of two
matrices. One of these matrices arises from the transition between
meson $B$ and one of the final mesons (form factor). The other
matrix is created by the residual end state due to the vacuum
(decay constant). In the studied decay, both matrices are for
pseudoscalar (spin 0), so the form factor becomes $\langle
B^-\rightarrow D^0\rangle$, and the decay constant takes form
$\langle 0\rightarrow D_s^-\rangle$. The main purpose of obtaining
hadron matrix elements is to estimate the amplitude. Then the
decay rate, branching ratio and $CP$ violation are obtained from
it. The branching fraction is obtained to be
$\mathcal{B}(B^-\rightarrow D_s^-D^0)=(9.33\pm1.17)\times10^{-3}$
at $\mu=m_b/2$ scale. This value is well compatible with the value
of $\mathcal{B}(B^-\rightarrow
D_s^-D^0)=(9.00\pm0.90)\times10^{-3}$ that reported by PDG(2020)
Avg. \cite{PDG1}. A value, $(-0.35\pm0.03)\%$, comparable to the
experimental result of LHCb \cite{LHCb2}, $(-0.4\pm0.5\pm0.5)\%$,
is obtained for $CP$ violation.

\section{Branching fraction and $CP$ asymmetry in $B^-\rightarrow D_s^-D^0$ decay}

\begin{figure}[t]
\centering \scalebox{0.7}{\includegraphics{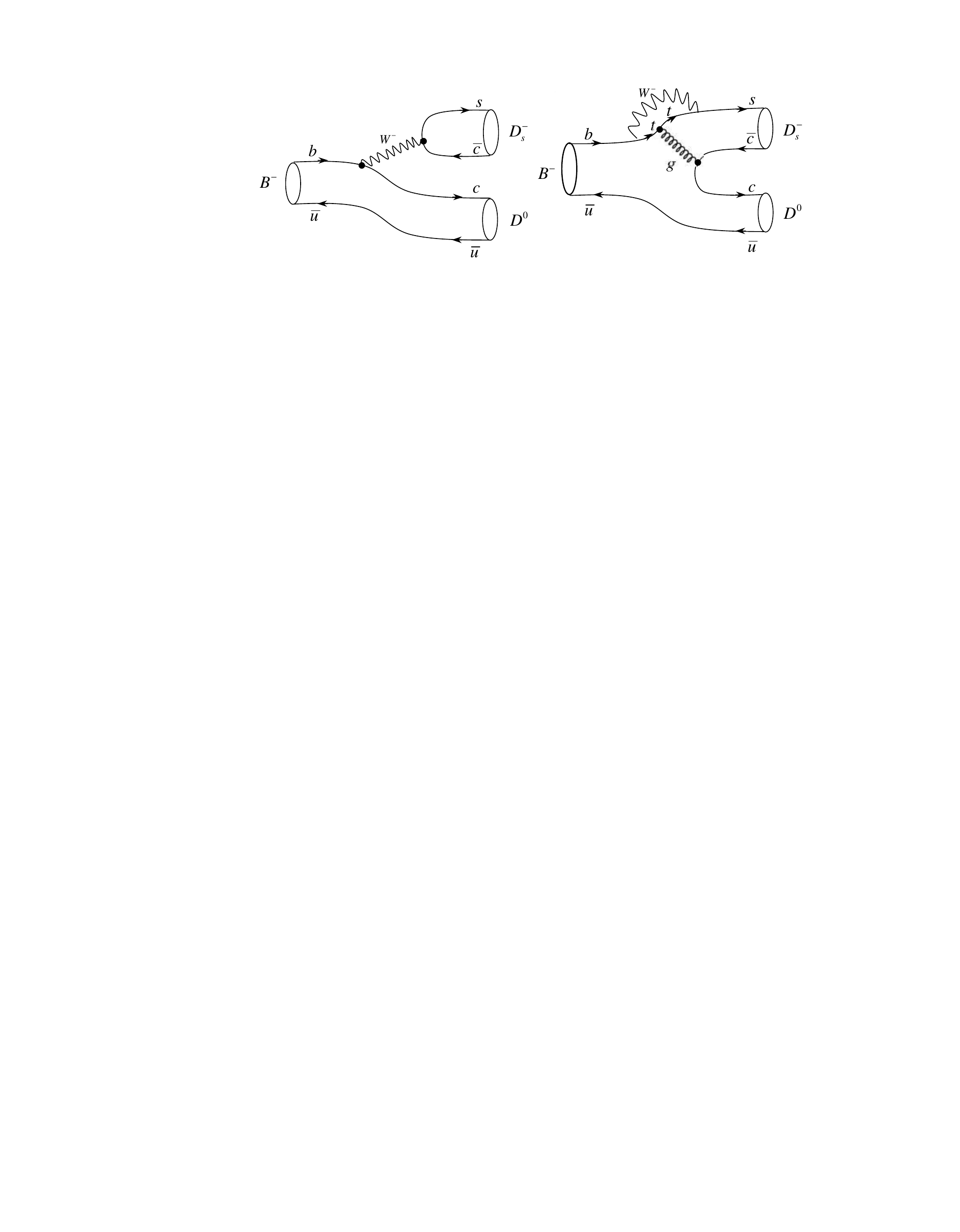}}
\caption{\label{fig1}Feynman diagrams contributing to
$B^-\rightarrow D_s^-D^0$ decay.}
\end{figure}

In this section, we calculate the $CP$ asymmetry and the branching
ratio for comparison with experimentally measured results. In the
study of $CP$ violation in $B$ decays it turned out to be useful
to make a classification of $CP$ violating effects that is more
transparent than the division into the indirect and direct $CP$
violation. Generally, complex phases may enter the
particle-antiparticle mixing and in the decay processes through
the complex elements of CKM matrix. As the phases in mixing and
decay are convention dependent, the $CP$ violating effects depend
only on the differences of these phases. Three types of $CP$
violation are: $CP$ violation in mixing, $CP$ violation in decay,
$CP$ violation in the
interference of mixing and decay \cite{A.J.B}.\\
In this study, the $CP$ violation in decay may occur for
$B^-\rightarrow D_s^-D^0$ decay. $CP$ violation in decay may be
zero due to very close values of particle and antiparticle decay
rates. In order for this symmetry not to be zero, two different
contributions in the amplitude with the strong $(\delta_i)$ and
weak $(\phi_i)$ phases are needed. These could be for instance two
tree diagrams, two penguin diagrams or just a diagram of each of
them. Considering the process of $B^-\rightarrow D_s^-D^0$ decay
(and $B^+\rightarrow D_s^+\bar{D}^0$ decay), this decay mode can
proceed through two different elementary amplitudes $\mathcal
{A}_1$ and $\mathcal {A}_2$, this means that the decay can proceed
by two different paths: tree ($\mathcal {A}_1$) and penguin
($\mathcal {A}_2$) diagrams, for the total decay amplitude we can
write \cite{M.S.S}: {\setlength\arraycolsep{.75pt}
\begin{eqnarray}\label{eq1}
\mathcal {A}(B^- \rightarrow D_s^-D^0)&=&|\mathcal {A}_1|
e^{i\delta_1}e^{i\phi_1} +|\mathcal
{A}_2|e^{i\delta_2}e^{i\phi_2},
\end{eqnarray}}
where $|\mathcal {A}_1|$ and $|\mathcal {A}_2|$ represent
$|\mathcal {A}_1(B^-\rightarrow D_s^-D^0)|$ and $|\mathcal
{A}_2(B^-\rightarrow D_s^-D^0)|$. To obtain the antiparticle
amplitude $(B^+ \rightarrow D_s^+\bar{D}^0)$, in the Eq.
(\ref{eq1}), the weak phases $(\phi_i)$ becomes its complex
conjugate and the strong phases $(\delta_i)$
remain unchanged.\\
For the decay examined in this study, $B^-\rightarrow D_s^-D^0$
decay, the Feynman's main diagrams are the tree level diagram that
have the largest contribution in the amplitude and the penguin
level diagram, which is significantly smaller than the tree level
diagram. The Feynman diagrams of $B^-\rightarrow D_s^-D^0$ decay
are shown in Fig. \ref{fig1}
 and then the decay amplitude reads
{\setlength\arraycolsep{.75pt}
\begin{eqnarray}\label{eq7}
\mathcal{A}(B^-\rightarrow
D_s^-D^0)&=&\frac{iG_{F}}{\sqrt{2}}f_{D_s} F_0^{B \rightarrow
D}(m_{D_s}^2)\Big(V_{cb}V_{cs}^*a_1 - V_{tb}V_{ts}^*(a_4 + a_{10}
+ \xi(a_6 + a_8))\Big),
\end{eqnarray}}
where the tree and penguin level amplitudes are as follows,
respectively{\setlength\arraycolsep{.75pt}
\begin{eqnarray}
\mathcal{A}_1 (B^-\rightarrow
D_s^-D^0)&=&\frac{iG_{F}}{\sqrt{2}}f_{D_s} F_0^{B \rightarrow
D}(m_{D_s}^2)V_{cb}V_{cs}^*a_1,
\end{eqnarray}}
and {\setlength\arraycolsep{.75pt}
\begin{eqnarray}
\mathcal{A}_2 (B^-\rightarrow
D_s^-D^0)&=&\frac{iG_{F}}{\sqrt{2}}f_{D_s} F_0^{B \rightarrow
D}(m_{D_s}^2)V_{tb}V_{ts}^*\Big(a_4 + a_{10} + \xi(a_6 +
a_8)\Big),
\end{eqnarray}}
$\xi$ depends on properties of the final-state mesons involved and
is defined as \cite{L.X.1}
\begin{eqnarray}
\xi=\frac{2m_{D_s}^2}{(m_b-m_c)(m_c+m_s)}
\end{eqnarray}
 To calculate the form factor $F_0$ we take the form \cite{D.Me}
\begin{eqnarray}
f(q^2)=\frac{f(0)}{1-\sigma_1q^2/m_P^2+\sigma_2q^4/m_P^4},
\end{eqnarray}
here $q^2=P^2_{B^-}-P^2_{D^0}=P^2_{D_s}$. The value of the
parameter $m_P$ (pole mass), which is equal to the lowest
resonance mass, is fixed to its physical value for proper choice
of the quark-model parameters and for the reliability of the
calculations. With this description the $m_P$ is the $m_{B_c}$ for
$B\rightarrow D$ transition. The values of $f(0)$, $\sigma_1$ and
$\sigma_2$ are as follows \cite{D.Me}
\begin{eqnarray}
F_0^{B\rightarrow D}:\quad f(0)=0.67,\quad \sigma_1=0.78,\quad
\sigma_2=0.
\end{eqnarray}
The quantities $a_i$ (i = 1, \ldots, 10) are the following
combinations of the effective Wilson coefficients
\begin{eqnarray}
a_{2i-1}=c_{2i-1}+\frac{1}{3}c_{2i},\quad
a_{2i}=c_{2i}+\frac{1}{3}c_{2i-1},\quad i=1,2,3,4,5.
\end{eqnarray}
The Wilson coefficients, $c_i$, in the effective weak Hamiltonian
have been reliably evaluated by the next-to-leading logarithmic
order. To proceed, we use the following numerical values at three
different choices of $\mu$ scale, which have been obtained in the
NDR scheme and are shown in Tab. \ref{tab2}.
\begin{table}[t]
\centering\caption{\label{tab2}  Wilson coefficients $c_i$ in the
NDR scheme ($\alpha=1/129$) \cite{M.Be1}.}
\begin{tabular}{|c||c|c|c|}
  \hline
   NLO    & $\mu=m_b/2$ & $\mu=m_b$ & $\mu=2m_b$\\ \hline\hline
 $c_1$ & 1.137 & 1.081 & 1.045 \\ \hline
 $c_2$ & -0.295 & -0.190 & -0.113 \\ \hline
 $c_3$ & 0.021 & 0.014 & 0.009 \\ \hline
 $c_4$ & -0.051 & -0.036 & -0.025 \\ \hline
 $c_5$ & 0.010 & 0.009 & 0.007 \\ \hline
 $c_6$ & -0.065 & -0.042 & -0.027 \\ \hline
 $c_7/\alpha$ & -0.024 & -0.011 & 0.011 \\ \hline
 $c_8/\alpha$ & 0.096 & 0.060 & 0.039 \\ \hline
 $c_9/\alpha$ & -1.325 & -1.254 & -1.195 \\ \hline
 $c_{10}/\alpha$ & 0.331 & 0.223 & 0.144 \\ \hline
\end{tabular}
\end{table}
The meson and quark masses and decay constants needed in our
calculations are taken as (in units of MeV)
\cite{PDG1}{\setlength\arraycolsep{.75pt}
\begin{eqnarray}\label{eq9}
m_{B_c}&=&6274.9\pm0.8,\quad m_{D^-_s}=1968.35\pm0.07,\quad
m_{B\pm}=5279.34\pm0.12,\quad m_{D^0}=1864.84\pm0.05,\nonumber\\
m_b&=&4180^{+40}_{-30},\quad m_s=93^{+11}_{-5},\quad
m_c=1270\pm20,\quad f_{D_s}=241\pm3 \cite{E.Fo1}.\quad
\end{eqnarray}}
The decay rates corresponding to the $\mathcal{A}(B^-\rightarrow
D_s^-D^0)$ and $\mathcal{A}({B}^+\rightarrow D_s^+\bar{D}^0)$
amplitudes can be then written as \cite{M.Ar1}
{\setlength\arraycolsep{.75pt}
\begin{eqnarray}\label{eq2}
\Gamma(B^-\rightarrow D_s^-D^0)&=&\Big(|\mathcal {A}_1|
e^{i(\delta_1 + \phi_1)}+|\mathcal {A}_2|e^{i(\delta_2 + \phi_2)}\Big)^2,\nonumber\\
\Gamma({B}^+\rightarrow D_s^+\bar{D}^0)&=&\Big(|\mathcal {A}_1|
e^{i(\delta_1 - \phi_1)}+|\mathcal {A}_2|e^{i(\delta_2 -
\phi_2)}\Big)^2.
\end{eqnarray}}
The branching fraction for the $B^-\rightarrow D_s^-D^0$ decay is
given by
\begin{eqnarray}
\mathcal{B}(B^-\rightarrow D_s^-D^0)=\frac{\Gamma(B^-\rightarrow
D_s^-D^0)}{\Gamma_{tot}},
\end{eqnarray}
where the $\Gamma_{tot}$ for charged $B$ meson is
$(4.02\pm0.01)\times10^{-13}$ GeV. By dividing the difference
between these two decay rates by their sum, the $CP$ asymmetry in
decay rates is given by \cite{I.Be.1}
{\setlength\arraycolsep{.75pt}
\begin{eqnarray}\label{eq12}
\mathcal{A}_{CP}&=&\frac{\Gamma(B^-\rightarrow
D^-_sD^0)-\Gamma(B^+\rightarrow
D^+_s\bar{D}^0)}{\Gamma(B^-\rightarrow
D^-_sD^0)+\Gamma(B^+\rightarrow D^+_s\bar{D}^0)}\nonumber\\
&=&\frac {2|\mathcal{A}_2/\mathcal
{A}_1|\sin(\delta_1-\delta_2)\sin(\phi_1-\phi_2)}{1+|\mathcal
{A}_2/\mathcal{A}_1|^2+2|\mathcal
{A}_2/\mathcal{A}_1|\cos(\delta_1-\delta_2)\cos(\phi_1-\phi_2)}.
\end{eqnarray}}
It should be noted that the numerical value of the difference
between the tree and penguin amplitudes, $\mathcal{A}_1
(B^-\rightarrow D_s^-D^0)-\mathcal{A}_2 (B^-\rightarrow
D_s^-D^0)$, is obtained as a complex number. The strong phase
$\delta_1-\delta_2$ arises from the ratio of the imaginary part to
the real part. In fact, the argument of the difference of two
amplitudes gives the strong phase, the estimated
value is $\delta_1-\delta_2=89.94^\circ$.\\
As mentioned in the introduction section, the weak phase $\phi_1$
is calculated from the argument of the complex element of $V_{ts}$
in the CKM matrix, the result is achieved to be
$\phi_1=1.10^\circ$. Considering that the element $V_{ub}$ in the
CKM matrix is a real number and the weak phase $\phi_2$ comes from
the argument of this element, so we set $\phi_2=0$.\\
In the Standard Model, the
Cabibbo-Kobayashi-Maskawa (CKM) quark-mixing matrix is a unitary
matrix, in which an expansion is introduced by the small value of
$\lambda$. The CKM matrix at order $\lambda^5$ can be
parameterized as \cite{HFLAV2}
\begin{eqnarray}
V=\left(%
\begin{array}{ccc}
  1-1/2\lambda^2-1/8\lambda^4 & \lambda & A\lambda^3(\rho-i\eta) \\
  -\lambda+1/2A^2\lambda^5[1-2(\rho+i\eta)] & 1-1/2\lambda^2-1/8\lambda^4(1+4A^2) & A\lambda^2 \\
  A\lambda^3[1-(1-1/2\lambda^2)(\rho+i\eta)] & -A\lambda^2+1/2A\lambda^4[1-2(\rho+i\eta)] & 1-1/2A^2\lambda^4 \\
\end{array}%
\right)
\end{eqnarray}
Recent Particle Data Group (PDG) average values for the
Wolfenstein parameters are \cite{PDG1}
\begin{eqnarray}
\lambda=0.22650\pm0.00048,\quad A=0.790^{+0.017}_{-0.012},\quad
\bar{\rho}=0.141^{+0.016}_{-0.017},\quad \bar{\eta}=0.357\pm0.011,
\end{eqnarray}
where $\bar{\rho}=\rho(1-1/2\lambda^2)$ and
$\bar{\eta}=\eta(1-1/2\lambda^2)$ \cite{M.Ba.1}. The CKM matrix
elements used in this work are obtained by the above calculations
as follows (in units of $10^{-3}$){\setlength\arraycolsep{.75pt}
\begin{eqnarray}\label{eq12}
 V_{cb}&=&40.529,\quad  V_{cs}=973.198,\nonumber\\
 V_{tb}&=&999.179,\quad V_{ts}=-39.790-0.762i.
\end{eqnarray}}

\section{Numerical results and conclusion}

The numerical results of the $CP$ violation and $CP$-averaged
branching ratio for $B^-\rightarrow D_s^-D^0$ is presented in Tab.
\ref{tab3}.
\begin{table}[t]
\centering\caption{\label{tab3} The numerical results of $CP$
violation and $CP$-averaged branching ratio for $B^-\rightarrow
D_s^-D^0$ decay at three different choices of $\mu$ scale.}
\begin{tabular}{|c||c|c|c|c|}
  \hline
$B^-\rightarrow D_s^-D^0$ & $\mu=m_b/2$ & $\mu=m_b$ & $\mu=2m_b$ &  Exp. \\
\hline\hline
 $\mathcal{A}_{CP} (\%)$ & $-0.96\pm0.09$ & $-0.35\pm0.03$ & $-0.11\pm0.01$ & $-0.4\pm0.5\pm0.5$ \cite{LHCb1}\\
 \hline
 $\mathcal{B}(\times10^{-3})$ & $9.33\pm1.17$ & $10.12\pm1.31$ & $10.75\pm1.41$ & $9.00\pm0.90$ \cite{PDG1}\\
\hline
\end{tabular}
\end{table}
In this paper, we have analyzed the decay of $B$ meson into two
pseudoscalar mesons. We have drawn Feynman diagrams completely for
the $B^-\rightarrow D_s^-D^0$ decay based on the standard model.
This decay can violate $CP$ symmetry. In general, $CP$ asymmetry
can be calculated from the difference between the particle and the
antiparticles decay rates relative to their sum and it becomes
non-zero for the mentioned decay. Here we have obtained the $CP$
violation by calculating the amplitude of the current-current tree
and penguin diagrams that are considered separately and using the
strong ($\delta_i$) and weak ($\phi_i$) phases. The weak and
strong phases have been obtained by complex elements of CKM matrix
and from current-current tree and penguin amplitude differences,
respectively. We have estimated the $CP$ violation as
$\mathcal{A}_{CP}(B^-\rightarrow D_s^-D^0)=(-0.35\pm0.03)\%$.
Also, from the sum of the amplitudes, we have calculated the total
amplitude and obtained comparable results with experimental values
for the branching ratio as: $\mathcal{B}(B^-\rightarrow
D_s^-D^0)=(9.33\pm1.17)\times10^{-3}$
at $\mu=m_b/2$ scale.\\
Theoretical uncertainties in our calculations are due to the
uncertainties in the form factors, decay constants, meson masses
and the uncertainties of the input parameters in CKM elements.

\end{document}